# Analytical solution of option pricing for two Stocks by time fractional ordered Black Scholes partial differential equation

Dr. Kamran Zakaria, Muhammad Saeed Hafeez

Department of Mathematics, NED University of Engineering and Technology, Karachi

**Abstract**

Time fractional order Black scholes partial differential equation for risk free option pricing in financial market yields the better prediction in financial market of the country. In this paper the modified form of Black Schole equation including two stocks is used for evaluations. Samudu Transform approach is utilized for calculating analytical solution . Solution of the equation has been found in form of convergent infinite series.

**Keywords:**

Options, Samudu Transforms, Caputo Derivative, the time fractional order partial differential equation, Black Scholes Equation.

**Introduction:**

This paper is the extended version of the paper presented in 3rd International Conference on Computing, Mathematics and Engineering Technologies (iCoMET) 2020 as given in the reference [1].

Option pricing is the worldwide growing field of financial mathematics to solve financial market pricing problems. It is a not very old discipline consists of techniques to predict the prices of assets in financial market. The Black Scholes mathematical model with aid of computer science provides the tool to solve complex financial markets, stock exchange and industrial problems. under discussion model is the benchmark imitations in finance, and it is the first mathematical models which predicts the pricing of options (both call and put) and the implied volatility.

The Maximum payoff is always the wish of successful businessman and it is possible only when the risk is minimized and gain is maximized. Specifically, the stock exchange is the good example for option pricing where the prices of shares are highly uncertain and unpredictable. But the risk free prices may be obtained by using the famous Black Scholes partial differential equation.

Price and share option valuation of options have been a comer stone in financial markets. Financial study of financial derivatives is one of two most growing areas in the corporate business finance . The mathematical models are servings to measure the variation, predict and forecast the behavior of financial markets. The Fractional calculus presents a highly endorsement and latest tool in business finance. Time fractional ordered Business Financial models are expressed in form fractional ordered stochastic PDE to maintain better accuracy and tolerance, to control variations and investigate random -ness in financial markets.

Black-Scholes model is the prominent and well known models to evaluate the option prices containing the brief literature review of its empirical developments theme of beginning with Black and Scholes (1973) experimental examinations close predisposition inside the Black-Scholes model as far as moneyness and development.

Studies have been growing additionally noted instability inclination operating at a profit Black Scholes model (1973) utilizing S&P 500 choice list information 1966-1969 recommend the fluctuation that appertain the choice delivers a cost among the framework’s cost and market cost. Black and Scholes (1973) propose proof, instability isn't fixed. Galai (1977) affirm B-S model that the supposition of verifiable momentary instability need to be loose. Their observational outcomes are likewise steady with the consequences of Geske in the year 1979.MacBeth and Merville (1980) look at the Black-Scholes model against the steady flexibility of difference (CEV) model, which expect unpredictability changes when the stock costs changes. MacBeth and Merville (1980) found that the unpredictability of the hidden stock minimized the risk as the stock value rises.

Beckers (1980) tried the Black-Scholes suspicion that the chronicled quick unpredictability of the fundamental stock is a component of the stock value, utilizing S&P 500 record alternatives 1972-1977. Beckers (1980) finds the hidden stock is a converse capacity of the stock cost.

Geske and Roll (1984) show that at a unique time both in-the-cash and out-of-the-cash choices contain instability inclination. Geske and Roll (1984) finish up, time and cash predisposition might be identified with inappropriate limit conditions, where as the unpredictability inclination issue might be the consequence of measurable mistakes in estimation.

Rubinstein (1994) shows that the inferred instability for S&P 500 file choices applies abundance kurtosis. Shimko (1993) exhibits that inferred conveyances of S&P 500 record are contrarily slanted and leptokurtic. Jackwerth and Rubinstein (1996) show the dispersion of the S&P 500 preceding 1987 apply lognormal appropriations, however since have disintegrated to look like leptokurtosis and negative skewness. A few examinations try to expand the tail properties of the lognormal conveyance by consolidating a hop dissemination process or stochastic instability.

Das and Sundaram (1999) show hop dispersion and stochastic instability moderate yet don't take out unpredictability predisposition. Das and Sundaram (1999) recognize hop dispersion and stochastic instability forms don't produce skewness and extra kurtosis looked like actually.

Buraschi and Jackwerth (2001) create measurable tests dependent on prompt model and stochastic models utilizing S&P 500 list alternatives information from 1986-1995. Buraschi and Jackwerth (2001) close the information is progressively steady with models that contain extra hazard factors, for example, stochastic unpredictability and hop dissemination.

Yang (2006) finds suggested volatilities used to esteem trade exchanged call alternatives on the ASX 200 Index are fair-minded and better than chronicled immediate unpredictability in guaging future acknowledged instability. Yang (2006) finds inferred volatilities used to esteem trade exchanged call alternatives on the ASX 2000 Index are fair and better than authentic immediate instability in determining future acknowledged unpredictability. Writing proposes the Black-Scholes model may undervalue alternatives in light of the fact that the tail properties of the hidden lognormal dissemination are excessively little.

In 2016, H. Zhang, F. Liu I. Turner, Q. Yang solved time fractional Black–Scholes model governing equation for European options numerically. In 2019, D. Prathumwan and K. Trachoo solved Black Sholes equation by the method of Laplace Homotopy Perturbation Method for two asset option pricing [1].

In this paper, the technique of Samudu transform method is used to demonstrate the analytical solution of 2-Dimensional, Time Fractional-ordered BS-Model, consists of two different assets in Liouville- Caputo Fractional derivative form for the European call options. The Sammudu Transform provides the value of put option in form of explicit solution in convergent series. In Lapace perturbation method, first step the method of Laplace transform is applied then homotopy method is applied. The solution is found by the certain hectic work but in the method of Samudu Transform, the solution may be found without such a hectic working. The solution is similar to the solution obtained by the method of Laplace Homotopy method. The method for solving two assets BS financial model is described in next section.

## Methodology

Consider P.D.E.

$$\frac{\partial^{\propto}}{\partial t^{\propto}} \emptyset\,(x,y,t) + \ \llcorner\emptyset(x,y,t) + N\emptyset(x,y,t) = f(x,y,t) --------(i)$$

Where $n-1 <\propto\leq n;\qquad n \in N$

Subject to:

$\emptyset(x,y,0) = \emptyset_0\,(x,y)$

Where $L =$ Linear Differential operator

$N =$ Non -Linear Differential operator

Call Sumudu Transform for the caputo Fractional ordered Derivative of $\emptyset(x,y,t)$on because of equation (i)

$$S\left[\frac{\partial^{\propto}}{\partial t^{\propto}}\ \emptyset\ (x,y,t)\right] = s[\ \emptyset(x,y,t) - \ \llcorner\emptyset(x,y,t) - N\emptyset(x,y,t) + \ f(x,y,t) - - - - - - - - (ii)$$

$$S\left[\frac{\partial^{\propto}}{\partial t^{\propto}}\ \emptyset\ (x,y,t)\right] = u^{-\propto}s[\emptyset(x,y,t) - \sum_{k=0}^{n-1} u^{-\propto+k}\ \frac{\partial^{k}}{\partial t^{k}}\ \emptyset(x,y,0)$$

If $\propto < 1$, setting $n = 1$

$$S\left[\frac{\partial^{\propto}}{\partial t^{\propto}}\ \emptyset\ (x,y,t)\right] = u^{-\propto}s\ \emptyset(x,y,t) - u^{-\propto}\ \emptyset(x,y,0)$$

$$u^{-\propto}s\ \emptyset(x,y,t) - u^{-\propto}\ \emptyset(x,y,0) = s\ [f(x,y,t) - \ \llcorner\ \emptyset(x,y,t) - N\emptyset(x,y,t)$$

$$s\ \emptyset(x,y,t) = \emptyset(x,y,0) + u^{-\propto}s\ [f(x,y,t) - \ \llcorner\ \emptyset(x,y,t) - N\emptyset(x,y,t)$$

Apply Inverse Sumudu Transform.

$$s\ s^{-1}\emptyset(x,y,t) = s^{-1}\ \emptyset(x,y,0) + s^{-1}\{u^{\propto}s\ [f(x,y,t) - \ \llcorner\ \emptyset(x,y,t) - N\emptyset(x,y,t)]\}$$

Call Inverse property of Samudu Transform (S.T)

$$I^{\propto}\ [g(x,y,t)] = s^{-1}[u^{\propto}s(g(x,y,t)]$$

Apply Integral Property of S.T on equation (3)

$$\emptyset(x,y,t) = \emptyset_0(x,y) + I^{\propto}\ [f(x,y,t) - \ \llcorner\ \emptyset(x,y,t) - N\emptyset(x,y,t) - - - - - - (4)$$

Sumudu Transform Express the so$l^{n}$ of P.D.E in form of Infinite Convergent series as below. (1)

$$\emptyset(x,y,t) = \emptyset_0(x,y) + \sum_{n=1}^{\infty} \frac{g_n(x,y)t^{n\propto}}{\Gamma(1+n\propto)}$$

Where

$$\emptyset_0(x,y) = g_0(x,y) = g_0$$

$g_1 = f(x, y, t) - \llcorner(g_0) - N(g_0)]$

$g_2 = f(x, y, t) - \llcorner(g_1) - N(g_1)]$

$g_3 = f(x, y, t) - \llcorner(g_2) - N(g_2)]$

$\vdots \qquad \qquad \vdots$

$$g_n = f(x, y, t) - \llcorner\, g_{(n)} - N[g_{(n)}]$$

This rearch paper is the application of Sumudu Integral Transform to evaluate option price of two stocks Fractional order Black Sholes Model.

Consider below Fractional order European call option pricing P.D.E for two stocks.

$$\frac{\partial^{\propto} c}{\partial t^{\propto}} + \frac{\sigma_1^2}{2} s_1^2 \frac{\partial^2 c}{\partial s_1^2} + \frac{\sigma_2^2}{2} s_2^2 \frac{\partial^2 c}{\partial s_2^2} + r s_1 \frac{\partial c}{\partial s_1} + r s_1 \frac{\partial c}{\partial s_1} + r s_2 \frac{\partial c}{\partial s_2} + p\, s_1 s_1 \sigma_1 \sigma_2 \frac{\partial^2 c}{\partial s_1 \partial s_2} - rc = 0$$

Subject to pay-off to the investor.

$c(s_1, s_{2\,,} t) = \max\ (w_1\ s_1 + w s_2 - \ k\ , 0)$ ( for European Call Option)

$\mathrm{P}(s_1, s_{2\,,} t) = \max\ (w_1\ s_1 + w s_2 - \ k\ , 0)$ ( for American Put option)

Where

c = value of European Call Option

P = value of American Call Option

$s_1$ = Price of share of stock 1

$s_2$ = Price of share of stock 2

P= correlation coefficient between price of shares of stock 1 and stock 2

$\sigma_1$= Price volatility or S.D of stock 1

$\sigma_2$= Price volatility or S.D of stock 2

K = Strike price or Exercise price for call option

$w_1$= properties of investment on stock 1

$w_2$= properties of investment on stock 2

Equation (4) can be simplified by considering the substitution.

$$u = lns_1 - \left(r - \frac{1}{2}\sigma_1^2\right)t$$

$$u = lns_2 - \left(r - \frac{1}{2}\sigma_2^2\right)t$$

$$\frac{\partial c}{\partial s_1} = \frac{\partial c}{\partial u}\frac{\partial u}{\partial s_1} \qquad \frac{\partial c}{\partial s_2} = \frac{\partial c}{\partial v}\frac{\partial u}{\partial s_2}\text{c}$$

$$(1) \ \ldots\ldots\ldots \frac{\partial c}{\partial s_1} = \frac{1}{s_1}\frac{\partial c}{\partial u} \qquad \frac{\partial c}{\partial s_2} = \frac{1}{s_2}\frac{\partial c}{\partial v} \ldots\ldots\ldots (2)$$

$$\frac{\partial^2 c}{\partial s_1^2} = \frac{\partial}{\partial s_1}\left(\frac{\partial c}{\partial u}\right)$$

$$= \frac{1}{s_1} = \frac{\partial}{\partial s_1}\left(\frac{\partial c}{\partial u}\right) + \frac{\partial c}{\partial u}\frac{\partial}{\partial s_1}\left(\frac{1}{s_1}\right)$$

$$= \frac{1}{s_1} = \frac{\partial}{\partial u}\left(\frac{\partial c}{\partial u}\right)\frac{\partial u}{\partial s_1} - s_1^2\,\frac{\partial c}{\partial u}$$

$$= \frac{1}{s_1^2}\frac{\partial^2 c}{\partial u^2} - \frac{1}{s^2}\frac{\partial c}{\partial u}$$

$$(ii) ------ \frac{\partial^2 c}{\partial s_1^2} = \frac{1}{s_1^2} - \frac{\partial^2 c}{\partial u^2} - \frac{\partial c}{\partial u}$$

Similarly

$$\frac{\partial^2 c}{\partial s_2^2} = \frac{1}{s_2^2}\left[\frac{\partial^2 c}{\partial v^2} - \ \frac{\partial c}{\partial v}\right] ------ (iii)$$

$$\frac{\partial c}{\partial s_1} = \ \frac{1}{s_1}\,\frac{\partial c}{\partial u}$$

$$\frac{\partial c}{\partial s_2}\left(\frac{\partial c}{\partial s_1}\right) = \ \frac{\partial}{\partial s_2}\ \left(\frac{1}{s_1} - \frac{\partial c}{\partial u}\right)$$

$$= \ \frac{1}{s_1}\frac{\partial}{\partial s_2}\ \frac{\partial c}{\partial u}$$

$$= \ \frac{1}{s_1}\frac{\partial}{\partial v}\left(\ \frac{\partial c}{\partial u}\right)\frac{\partial v}{\partial s_2}$$

$$= \ \frac{1}{s_1}\frac{\partial^2 c}{\partial u\ \partial v}\ \frac{1}{s^2}$$

$$(iv) ------- \frac{\partial^2 c}{\partial s_1\ \partial s_2}\ \frac{1}{s_1\ s_2}\ \frac{\partial^2 c}{\partial u\ \partial v}$$

Substitute (i) (ii) , (iii) & (iv) in equation .....(4)

$$\left\{\begin{array}{l} \frac{\partial^{\propto} c}{\partial t^{\propto}} + \ \frac{s_1^2\ \sigma_1^2}{\partial s_1^2}\left(\frac{\partial^2 c}{\partial u^2} - \ \frac{\partial c}{\partial u}\right) + \frac{\sigma^2}{2}\frac{s_2^2}{s_2^2} = \ \left(\frac{\partial^2}{\partial v^2} - \frac{\partial c}{\partial v}\right) + \ \partial s_1 \,.\frac{1}{s_1}\ \frac{\partial c}{\partial u} \\ + r s_2 \,.\frac{1}{S_2}\,\frac{\partial c}{\partial v} + \ P\ s_1 s_2\ \sigma_1\sigma_2\ \frac{1}{S_1 S_2}\ \ \sigma_1\sigma_2\ \ \frac{1}{S_1\,S_2}\ \ \frac{\partial^2 c}{\partial u\ \partial v} - \ rc = 0 \end{array}\right.$$

$$\left\{\begin{array}{l} \frac{\partial^{\propto} c}{\partial t^{\propto}} + \ \frac{\sigma_1^2}{2}\ \frac{\partial^2 c}{\partial u^2} + \frac{\sigma_2^2}{2}\ \frac{\partial^2\,c}{\partial v^2} - \frac{\sigma_1^2}{2}\frac{\partial p}{\partial u} - \frac{\sigma_2^2}{2}\frac{\partial c}{\partial u} = \ \left(\frac{\partial^2}{\partial v^2} - \frac{\partial c}{\partial v}\right) \\ \\ \end{array}\right.$$

$$+ r\frac{\partial c}{\partial u} + r\frac{\partial c}{\partial v} + \; P\,\sigma_1\sigma_2 - \;\frac{\partial^2 c}{\partial u\,\partial v} - \;rc = 0$$

$$\begin{cases} \frac{\partial^\propto c}{\partial t^\propto} + \;\frac{\sigma_1^2}{2}\,\frac{\partial^2 c}{\partial u^2} + \frac{\sigma_2^2}{2}\,\frac{\partial^2 c}{\partial v^2} \;+\left(r\;\frac{\sigma_1^2}{2}\right)\frac{\partial c}{\partial u} + \;\left(r - \frac{\sigma_2^2}{2}\right)\,\frac{\partial c}{\partial v} \\ \qquad +p\,\sigma_1\sigma_2\,\frac{\partial^2 c}{\partial u\,\partial v} - \;rc = 0 \end{cases}$$

We have substitution

$$u = lns_1 - \left(r - \frac{1}{2}\,\sigma_1^2\right)\,t$$

$$v = lns_2 + \left(r - \frac{1}{2}\,\sqrt{2}^2\right)\,t$$

$$\frac{\partial c}{\partial t} = \frac{\partial c}{\partial u}\,\frac{\partial u}{\partial t}$$

$$\boxed{\frac{\partial c}{\partial t} = \;\frac{\partial c}{\partial u}\left(r - \;\frac{1}{2}\,\sigma_1^2\right)}$$

$$\frac{-\,\partial c}{\partial t} = \frac{\partial c}{\partial u}\,\left(r - \frac{1}{2}\,\sigma_1^2\right)$$

$$\frac{\partial c}{\partial t} = \frac{\partial c}{\partial v}\,\frac{\partial v}{\partial t}$$

$$\boxed{\frac{\partial p}{\partial t} = \;\frac{\partial c}{\partial v}\left(r - \;\frac{1}{2}\,\sigma_1^2\right)}$$ ----------(vii)

From (vi) & (vii) equation (v) becomes.

$$\frac{\partial^\propto c}{\partial t^\propto} + \;\frac{\sigma_1^2}{2}\,\frac{\partial^2 c}{\partial u^2} + \frac{\sigma_2^2}{2}\,\frac{\partial^2 c}{\partial v^2} - \frac{\partial c}{\partial t} + \frac{\partial c}{\partial t} + +p\,\sigma_1\sigma_2\,\frac{\partial^2 c}{\partial u\,\partial v} - \;rc = 0$$

$$\frac{\partial^{\propto} c}{\partial t^{\propto}} + \frac{\sigma_1^2}{2}\frac{\partial^2 c}{\partial u^2} + \frac{r_2^2}{2}\frac{\partial^2 c}{\partial v^2} + p\,\sigma_1\sigma_2\,\frac{\partial^2 c}{\partial u\,\partial v} - rc = 0$$ ----------(A)

Subject to:

$c(u, v, o) =$ max $(w_1\, e^u + w_2\, e^v, 0)$ ----------(viii)

Hence equation (viii) be the simplified European style ca;; option pricing model for two stocks

Fractional order Black-shole P.D.E

Apply sumudu transform on (viii)

$$s\left[\frac{\partial^{\propto} c}{\partial t^{\propto}}\right] = -\frac{\sigma_1^2}{2}\frac{\partial^2 c}{\partial u^2} + \frac{\sigma_2^2}{2}\frac{\partial^2 c}{\partial v^2} + p\,\sigma_1\sigma_2\,\frac{\partial^2 c}{\partial u\,\partial v} - rc = 0$$

$$u^{-\propto} S\, c\,(u, v, t) - u^{-\propto}\,(u, v, o) = -s\,[\frac{\sigma_1^2}{2}\frac{\partial^2 c}{\partial u^2} + \frac{\sigma_2^2}{2}\frac{\partial^2}{\partial v^2} c + p\,\sigma_1\sigma_2\,\frac{\partial^2 c}{\partial u\,\partial v} - rc = 0$$

$$S\,c\;(u, v, t) - c\,(u, v, o) = -u^{\propto} s\,[\frac{\sigma_1^2}{2}\frac{\partial^2 c}{\partial u^2} + \frac{\sigma_2^2}{2}\frac{\partial^2 c}{\partial v^2} + p\,\sigma_1\sigma_2\,\frac{\partial^2 c}{\partial u\,\partial v} - rc\,]$$

$$S\,c\;(u, v, t) = c\,(u, v, o) - u^{\propto} s\,[\frac{\sigma_1^2}{2}\frac{\partial^2 c}{\partial u^2} + \frac{\sigma_2^2}{2}\frac{\partial^2 c}{\partial v^2} + p\,\sigma_1\sigma_2\,\frac{\partial^2 c}{\partial u\,\partial v} - rc - - - - - \text{(ix)}$$

Apply inverse samudu transform on -----------(ix)

$$c\;(u, v, t) = s^{-1}\,c\,(u, v, \infty)\, s^{-1}\,[u^{\propto} s\,[\frac{\sigma_1^2}{2}\frac{\partial^2 c}{\partial u^2} + \frac{\sigma_2^2}{2}\frac{\partial^2 c}{\partial v^2} + p\,\sigma_1\sigma_2\,\frac{\partial^2 c}{\partial u\,\partial v} - rc$$

$$c\ \ (u,v,t) = c\ (u,v,o)\ s^{-1}\ [\ u^{\propto} s\ [\frac{\sigma_1^2}{2}\frac{\partial^2 c}{\partial u^2} + \frac{\sigma_2^2}{2}\frac{\partial^2\ c}{\partial v^2} + p\ \sigma_1\sigma_2\ \frac{\partial^2 c}{\partial u\ \partial v} -\ rc\ ------- \text{(x)}$$

By definition # 14

$$I^{\propto}\ g\ \ (x,y,t) = s^{-1}[u^{\propto} sg\ \ (x,y,t)]$$

Apply Integral property of Sumudu transform on -------------------- (x)

$$c\ \ (u,v,t) = c\ (u,v,o)\ \ I^{\propto}\ [\frac{\sigma_1^2}{2}\frac{\partial^2 c}{\partial u^2} + \frac{\sigma_2^2}{2}\frac{\partial^2\ c}{\partial v^2} + p\ \sigma_1\sigma_2\ \frac{\partial^2 c}{\partial u\ \partial v} -\ rc\ ------ \text{(ix)}$$

Sumudu transform expresses Sol$^{n}$ of P.D.E by using equation (xi) in form of infinite convergent series as below.

$$c_0\ \ (u,v,t) = c\ (u,v,o)\ \sum_{n=0}^{\infty}\frac{gn(x,y)t^{n\propto}}{\Gamma(1+n^{\propto})}$$

Where $c\ \ (u,v,t) = g_0\ (u,v) =\ g_0\ (say)$

$$c_{n+1}\ \ (u,v,t)\ = \sum_{n=0}^{\infty}\frac{gn\ t^{n\propto}}{\Gamma(1+n^{\propto})}$$

$$c\ \ (u,v,t) = c\ \ (u,v,o) + \sum_{n=1}^{\infty}\frac{gn(x,y)\ t^{n\propto}}{\Gamma(1+n^{\propto})}$$

be the European call option price solution at time t.

where

$$c\ \ (u,v,o) =\ g_0$$

$$g_1 = -\left[\frac{\sigma_1^2}{2}\frac{\partial^2 g_0}{\partial u^2} + \frac{\sigma_2^2}{2}\frac{\partial^2\ g_0}{\partial v^2} + p\ \sigma_1\sigma_2\ \frac{\partial^2 g_0}{\partial u\ \partial v} -\ rg_0\right]$$

$$g_2 = -\left[\frac{\sigma_1^2}{2}\frac{\partial^2 g_1}{\partial u^2} + \frac{\sigma_2^2}{2}\frac{\partial^2\ g_1}{\partial v^2} + p\ \sigma_1\sigma_2\ \frac{\partial^2 g_1}{\partial u\ \partial v} -\ rg_1\right]$$

$$g_3 = -\left[\frac{\sigma_1^2}{2}\frac{\partial^2 g_2}{\partial u^2} + \frac{\sigma_2^2}{2}\frac{\partial^2 g_2}{\partial v^2} + p\,\sigma_1\sigma_2\frac{\partial^2 g_2}{\partial u\,\partial v} - rg_2\right]$$

$$\vdots$$

$$g_{n+1} = -\left[\frac{\sigma_1^2}{2}\frac{\partial^2 g_n}{\partial u^2} + \frac{\sigma_2^2}{2}\frac{\partial^2 g_n}{\partial v^2} + p\,\sigma_1\sigma_2\frac{\partial^2 g_n}{\partial u\,\partial v} - rg_n\right]$$

## Illustrations of option Price For Two Stocks Fractional Ordered Black Sholes Model

**Illustrative example 1**: Option type: European call option, consider the following data.

$S_1$ = Price of stock 1 in dollors.

$S_2$ = Price of stock 2 in dollars

| | | | | | |
|---|---|---|---|---|---|
| $S_1$ | 20 | 40 | 70 | 100 | 150 |
| $S_2$ | 50 | 80 | 120 | 180 | 200 |

Initial Condition :

C$(s_1, s_2, t)$ =Max $(e^{s_1} + 2e^{s_2} - 80, 0)$

- Exercise price of stock 1 = Rs.80
- Exercise price of stock 2 = Rs.20
- Maximum Exercise price for = Rs.80
  Option type : European Call option.
- Month of Expiration or time for Exercise data = 8 months
- $\propto = 0.005$
- S.D of stock 1 $= \sigma_1 = 40$ %

- S.D of stock 2 = $\sigma_2$ = 25 %
- Proportion of stock 1 = $w_1 = 2$ & proportion of stock 2 =$w_2$=2
- Risk free rate of return = 8%
- Correlation coefficient between Stock 1 and stock 2= 75%

solving the above problem by using Matlab programming ,European call option price of the stocks is represented as below

$C(s_1, s_2, t)$ = *1.0512* $e^{s_1}$ *+ 2.1274* $e^{s_2}$ *- 86.942*

***CALL OPTION PRICES***

| | | Sock S1 | | | | |
|---|---|---|---|---|---|---|
| | | 20 | 40 | 70 | 100 | 150 |
| | 50 | 42.193 | 62.744 | 93.569 | 124.39 | 175.77 |
| | 80 | 107.13 | 127.68 | 158.5 | 189.33 | 240.7 |
| Stock S2 | 120 | 193.7 | 214.25 | 245.08 | 275.9 | 327.28 |
| | 180 | 323.57 | 344.12 | 374.94 | 405.77 | 457.15 |
| | 200 | 366.86 | 387.41 | 418.23 | 449.06 | 500.43 |

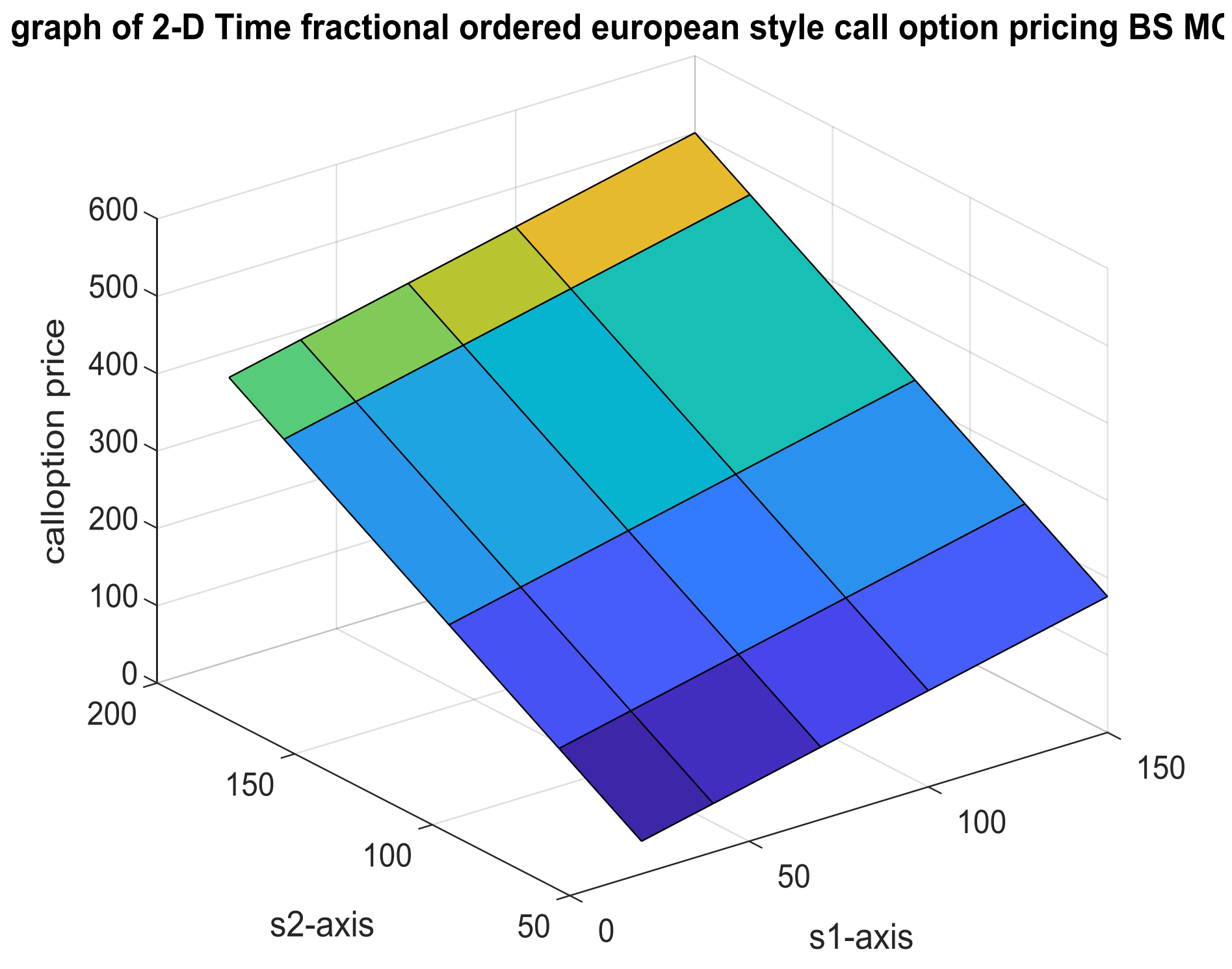

**llustrative example 2**: Option type : put option consider the following data.

$S_1$ = Price of stock 1 in dollors.

$S_2$ = Price of stock 2 in dollars

| | | | | | |
|---|---|---|---|---|---|
| $S_1$ | 20 | 40 | 70 | 100 | 150 |
| $S_2$ | 50 | 80 | 120 | 180 | 200 |

Initial Condition: $P(s_1, s_2, t) = Max((60 - 3\sin\pi s_1 - 5cos\pi s_2), 0)$

- Exercise price of stock 1 = Rs.60
- Exercise price of stock 2 = Rs.20
- Maximum Exercise price for = Rs.60
  Option type : put option.
- Month of Expiration or time for Exercise data = 2 months
- $\propto = 0.755$
- S.D of stock 1 = $\sigma_1$ = 45 %
- S.D of stock 2 = $\sigma_2$ = 85 %
- Proportion of stock 1 = $w_1 = 3$ & proportion of stock 2 =$w_2$=5

- Risk free rate of return = 03%
- Correlation coefficient between  Stock 1 and stock 2= 65%

solving the above problem by using  Matlab programming ,put option price of the stocks is represented as below

$$P(s_1, s_2, t) = 27.459\cos(3.1416 s_2) - 2.0559\cos(3.1416 s_1) - 1.4938\sin(3.1416 s_1) - 32.852\sin(3.1416 s_2) + 60.514$$

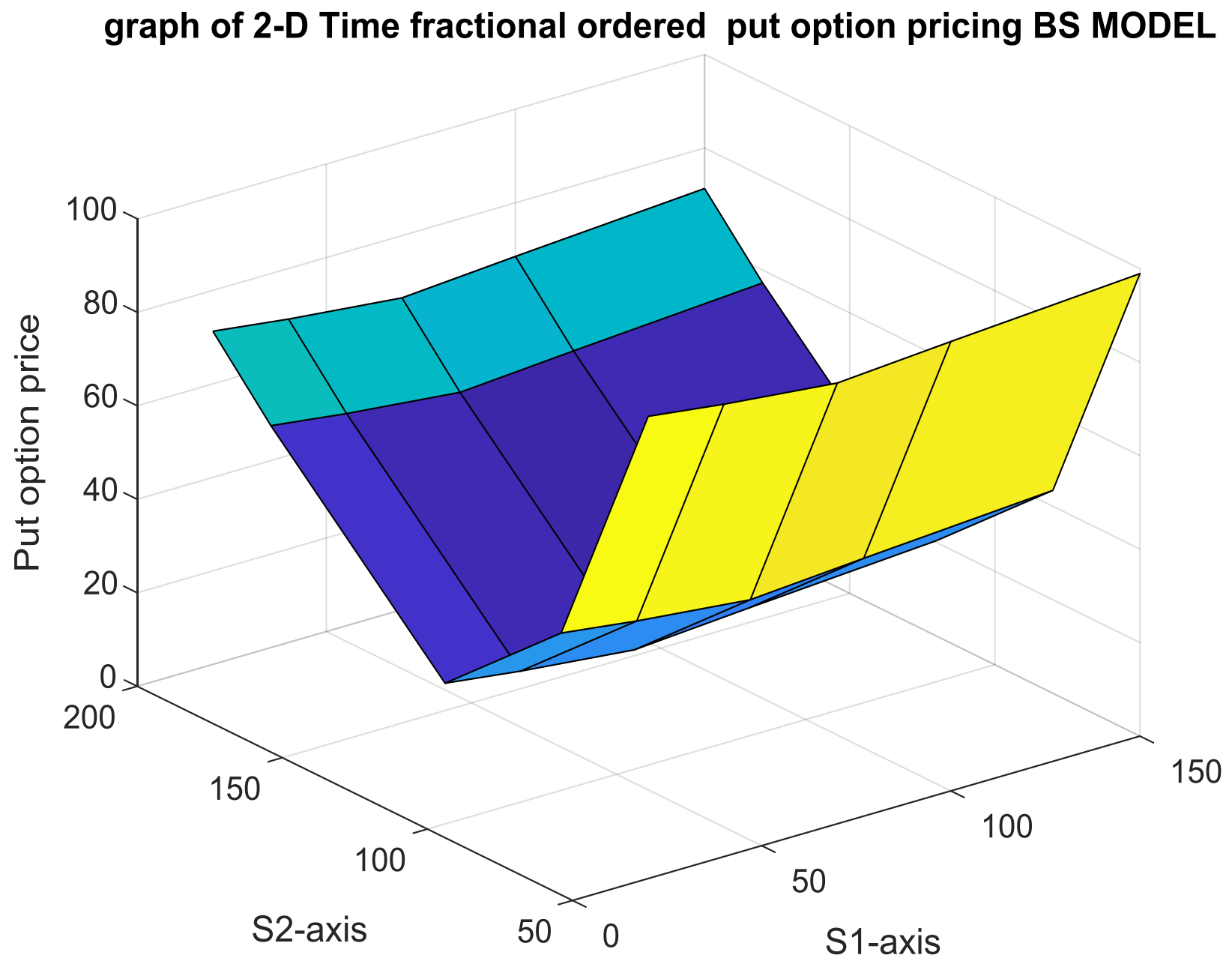

***PUT OPTION PRICES***

| | | Stock $S_1$ | | | | |
|---|---|---|---|---|---|---|
| | | 20 | 40 | 70 | 100 | 150 |
| Stock $S_2$ | 50 | 98.861 | 96.764 | 94.274 | 96.115 | 98.926 |
| | 80 | 43.341 | 41.244 | 38.754 | 40.595 | 43.406 |
| | 120 | 20.404 | 18.307 | 15.817 | 17.658 | 20.469 |
| | 180 | 57.17 | 55.072 | 52.583 | 54.424 | 57.235 |

| | | | | | |
|---|---|---|---|---|---|
| 200 | 71.267 | 69.17 | 66.68 | 68.521 | 71.332 |

**Illustrative example 3**: Option type : European call option consider the following data.

$S_1$ = Price of stock 1 in dollors.

$S_2$ = Price of stock 2 in dollars

| | | | | | |
|---|---|---|---|---|---|
| $S_1$ | 20 | 40 | 70 | 100 | 150 |
| $S_2$ | 50 | 80 | 120 | 180 | 200 |

Initial condtion: $C(s_1, s_2, t) = Max((2s_1^3 + 5s_2^2), 0)$

- Exercise price of stock 1 = Rs.60
- Exercise price of stock 2 = Rs.90
- Maximum Exercise price for = Rs.90
  Option type : put option.
- Month of Expiration or time for Exercise data = 2 months
- $\propto = .125$
- S.D of stock 1 = $\sigma_1$ = 40 %
- S.D of stock 2 = $\sigma_2$ = 65 %
- Proportion of stock 1 = $w_1 = 2$ & proportion of stock 2 =$w_2$=5
- Risk free rate of return = 07%
- Correlation coefficient between Stock 1 and stock 2= 85%
  solving the above problem by using matlab programming ,put option price of the stocks is represented as below

$$C(s_1, s_2, t) = 2.1517\, s_1^3 + 5.3794\, s_2^2 - 99.777$$

| | Stock S1 | | | | |
|---|---|---|---|---|---|
| | 20 | 40 | 70 | 100 | 150 |

| | | | | | | |
|---|---|---|---|---|---|---|
| Stock S2 | 50 | 40.648 | 88.334 | 142.39 | 185.17 | 242.47 |
| | 80 | 60.194 | 107.88 | 161.94 | 204.71 | 262.02 |
| | 120 | 78.847 | 126.53 | 180.59 | 223.36 | 280.67 |
| | 180 | 99.157 | 146.84 | 200.9 | 243.67 | 300.98 |
| | 200 | 104.71 | 152.39 | 206.45 | 249.22 | 306.53 |

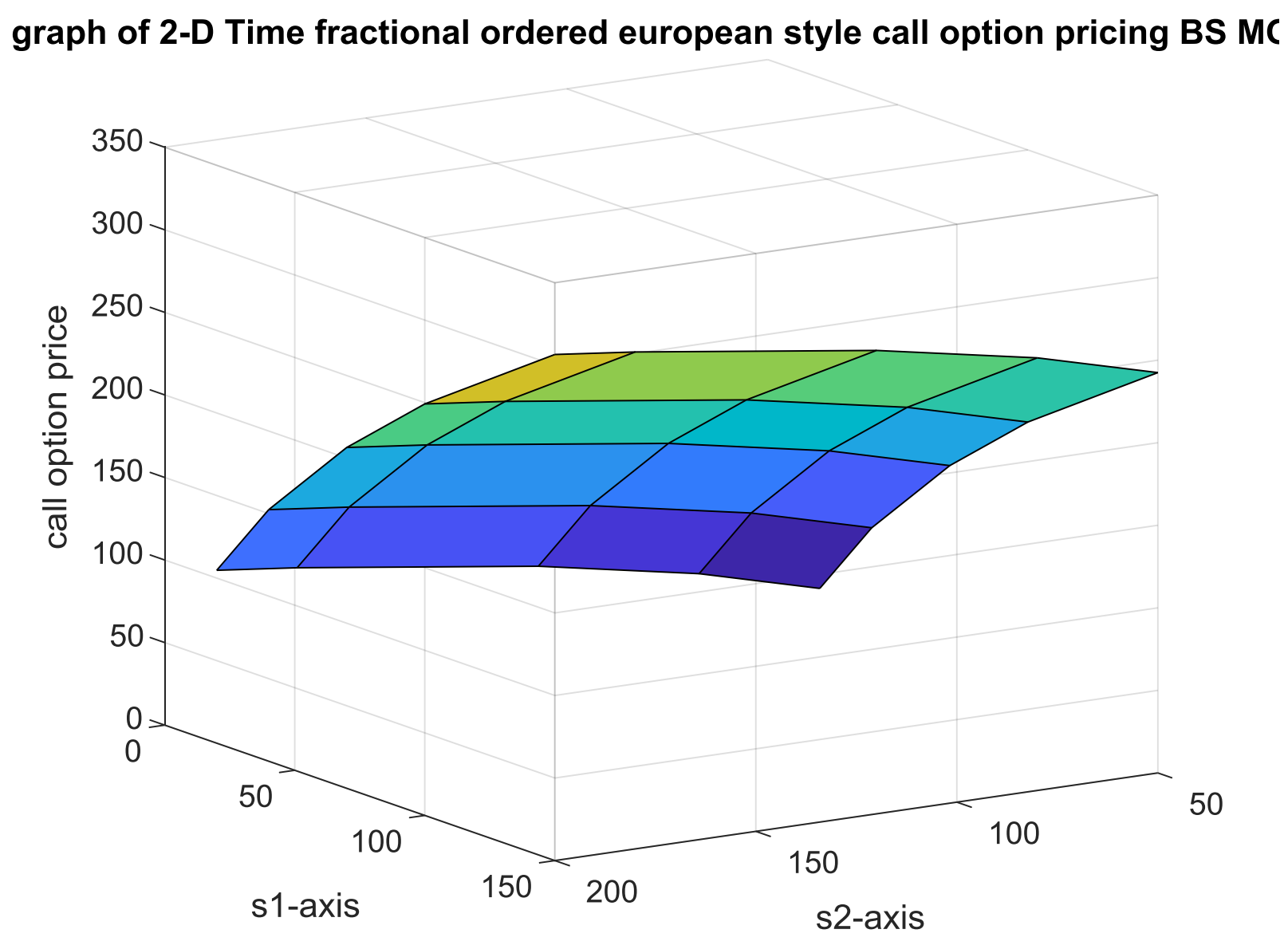

**<u>Illustrative example 4</u>**: Option type : European put option , consider the following data.

$S_1$ = Price of stock 1 in dollors.

$S_2$ = Price of stock 2 in dollars

| | | | | | |
|---|---|---|---|---|---|
| $S_1$ | 20 | 40 | 70 | 100 | 150 |
| $S_2$ | 50 | 80 | 120 | 180 | 200 |

Initial condtion: $C(s_1, s_2, t) = Max($2$(x^2$+$y^2)$ - ln(y)+ ln(x)-,0)

- Maximum Exercise price for = Rs. $2(x^2+y^2)$
  Option type : put option.
- Month of Expiration or time for Exercise data = 5 months
- $\propto = .125$
- S.D of stock $1 = \sigma_1 \quad = 40$ %
- S.D of stock $2 = \sigma_2 \quad = 20$ %
- Proportion of stock 1 = $w_1 = 1$ & proportion of stock 2 =$w_2$=1
- Risk free rate of return = 8%
- Correlation coefficient between Stock 1 and stock 2**= 75**%

  solving the above problem by using matlab programming ,put option price of the stocks is represented as below:

$$p(s_1, s_2, t) = 2.1735x^2 - 1.0868\ln(y) - 1.0868\ln(x) + 0.073569/x^3 + 0.14482/x^6 + 1.2248/x^9 + 2.1735y^2 + 0.047084/y^2 + 0.05932/y^6 + 0.32106/y^9 - 48.905$$

***Put -0ption prices***

| | | Stock $S_1$ | | | | |
|---|---|---|---|---|---|---|
| | | 20 | 40 | 70 | 100 | 150 |
| Stock $S_2$ | 50 | 28.452 | 46.868 | 60.826 | 68.965 | 77.391 |
| | 80 | 31.964 | 52.15 | 67.538 | 76.588 | 86.049 |
| | 120 | 34.994 | 56.708 | 73.328 | 83.164 | 93.519 |
| | 180 | 38.025 | 61.265 | 79.119 | 89.74 | 100.99 |
| | 200 | 38.812 | 62.45 | 80.623 | 91.449 | 102.93 |

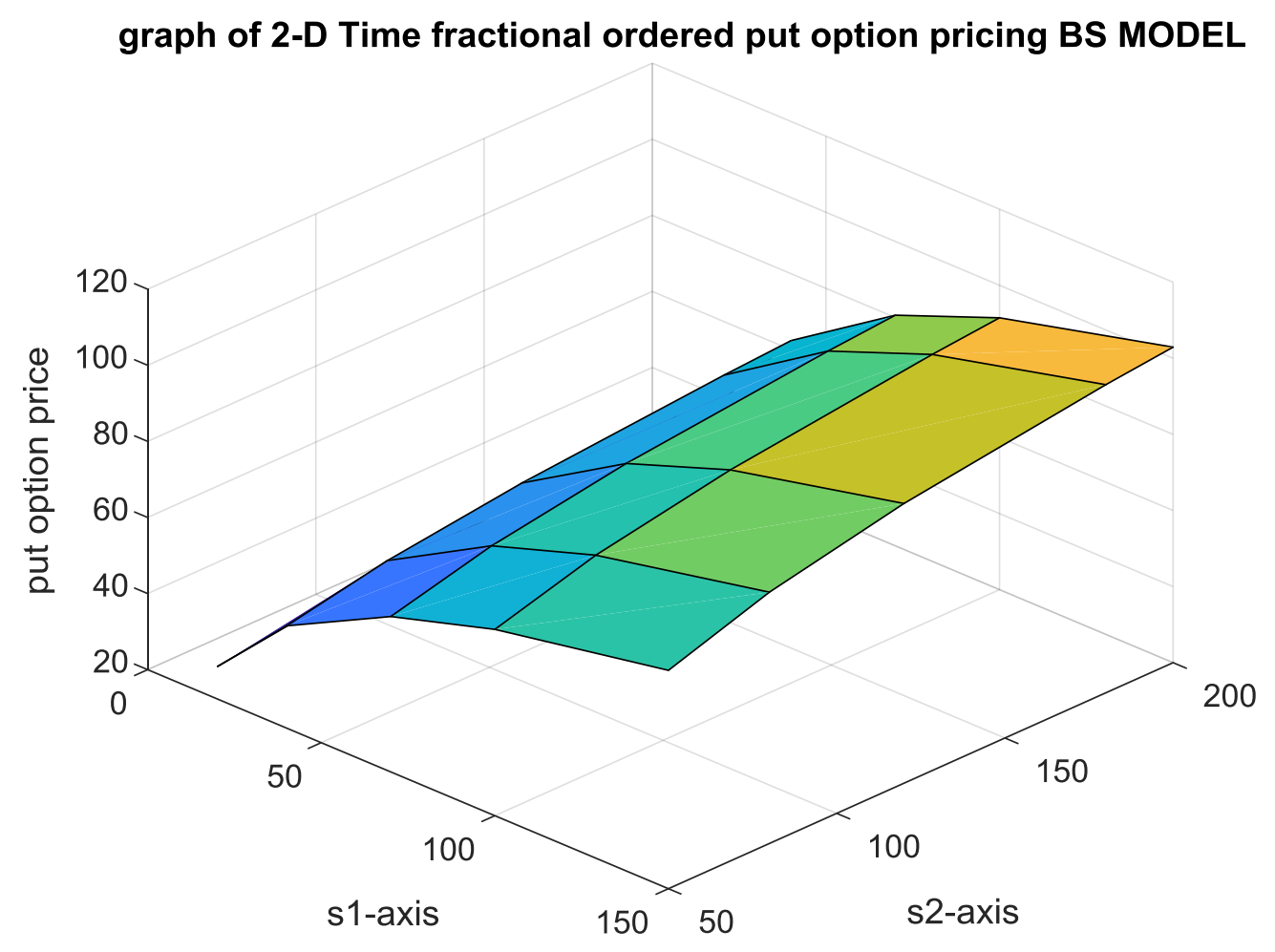

---

**<u>Illustrative example 5</u>:** Option type: put option , consider the following data.

$S_1$ = Price of stock 1 in dollors.

$S_2$ = Price of stock 2 in dollars

| | | | | | |
|---|---|---|---|---|---|
| $S_1$ | 20 | 40 | 70 | 100 | 150 |
| $S_2$ | 50 | 80 | 120 | 180 | 200 |

Initial condtion: $p(s_1, s_2, t) = Max($-5sin(x)-8y+5xy,0);x)-,0)

- Maximum Exercise price for = Rs. 25xy
  Option type : put option.
- Month of Expiration or time for Exercise data = 5 months
- $\propto = .125$
- S.D of stock $1 = \sigma_1 = 40\ \%$
- S.D of stock $2 = \sigma_2 = 20\ \%$
- Proportion of stock 1 = $w_1 = 1$ & proportion of stock 2 =$w_2$=1
- Risk free rate of return = 8%
- Correlation coefficient between Stock 1 and stock 2= **75%**

  solving the above problem by using matlab programming ,put option price of the stocks is represented as below:

$$P(s_1, s_2, t) = 5.4338xy - 0.18377\cos(x) - 5.4278\sin(x) - 8.6942y - 0.22071$$

***PUT-OPTION PRICES***

| | | Stock $S_1$ | | | | |
|---|---|---|---|---|---|---|
| | | 20 | 40 | 70 | 100 | 150 |
| Stock $S_2$ | 50 | 50.18 | 59.961 | 69.412 | 76.156 | 84.502 |
| | 80 | 58.571 | 68.353 | 77.803 | 84.548 | 92.894 |
| | 120 | 66.592 | 76.374 | 85.824 | 92.568 | 100.91 |
| | 180 | 75.335 | 85.117 | 94.568 | 101.31 | 109.66 |
| | 200 | 77.725 | 87.507 | 96.958 | 103.7 | 112.05 |

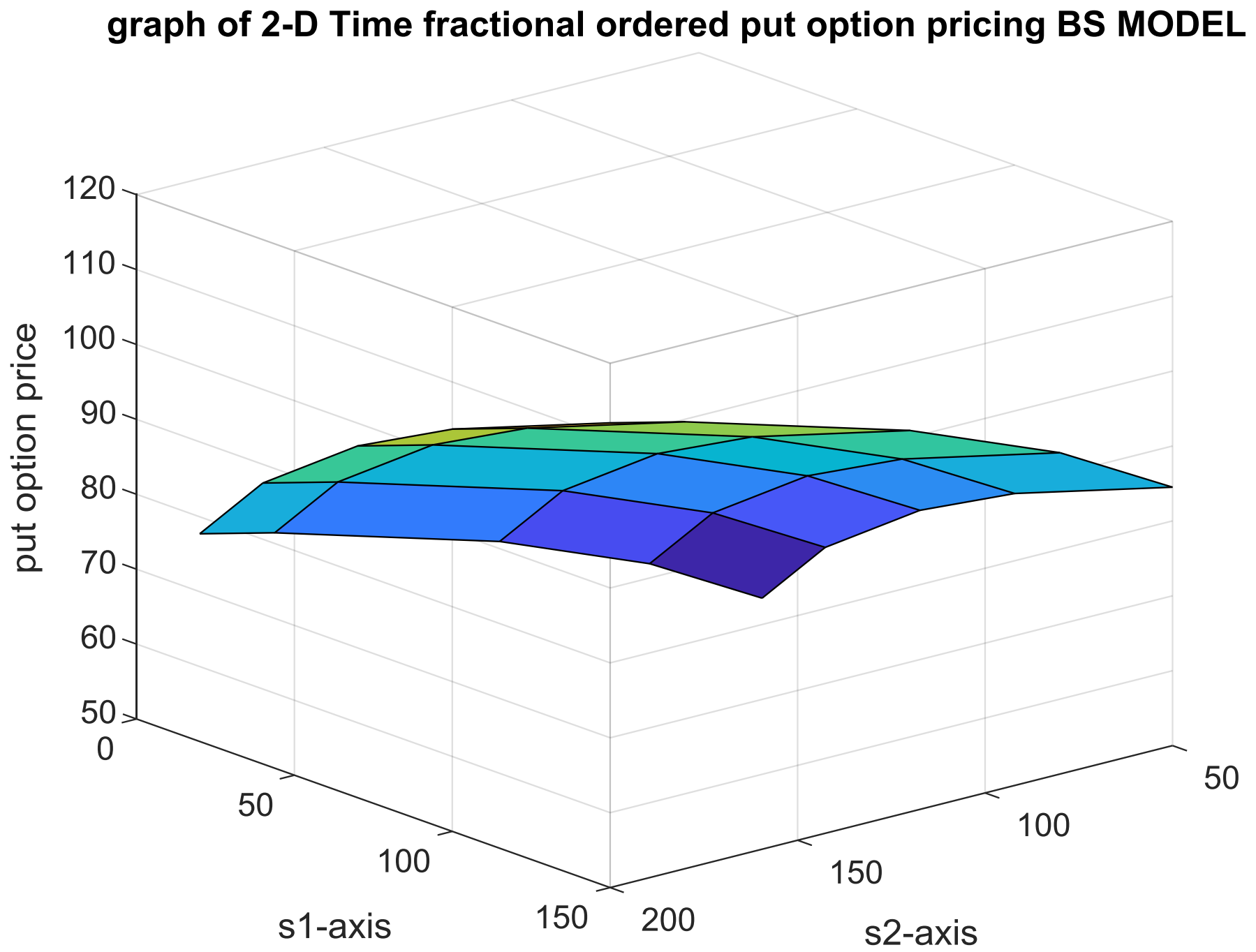

**Concluding Remarks** :

In this paper technique of Samudu Transforms and its derivatives and integral properties are applied to compute analytical solution of time fractional non – linear two dimensional BS PDE model in form of infinite series to evaluate put options of two stocks asset. Illustrative practical example is also presented to understand the reliability, efficiency, simplicity, and effectiveness of the purposed scheme. Samudu Transforms has many powerful and effective techniques to obtain analytic solution of any type time fractional PDE in least time with less computation.

---